# A specialized face-processing network consistent with the representational geometry of monkey face patches

Amirhossein Farzmahdi[1], Karim Rajaei[1], Masoud Ghodrati[2], Reza Ebrahimpour[3,1*], Seyed-Mahdi Khaligh-Razavi[4]

1 School of Cognitive Sciences (SCS), Institute for Research on Fundamental Sciences (IPM), Tehran, Iran
2 Department of Physiology, Monash University, Melbourne, VIC, Australia
3 Department of Computer Engineering, Shahid Rajaee Teacher Training University, Tehran, Iran
4 Computer Science and Artificial Intelligence Laboratory, Massachusetts Institute of Technology, USA

* E-mail: Corresponding ebrahimpour@ipm.ir

## 1 Abstract

Ample evidence suggests that face processing in human and non-human primates is performed differently compared with other objects. Converging reports, both physiologically and psychophysically, indicate that faces are processed in specialized neural networks in the brain – i.e. face patches in monkeys and the fusiform face area (FFA) in humans. We are all expert face-processing agents, and able to identify very subtle differences within the category of faces, despite substantial visual and featural similarities. Identification is performed rapidly and accurately after viewing a whole face, while significantly drops if some of the face configurations (e.g. inversion, misalignment) are manipulated or if partial views of faces are shown due to occlusion. This refers to a hotly-debated, yet highly-supported concept, known as holistic face processing. We built a hierarchical computational model of face-processing based on evidence from recent neuronal and behavioural studies on faces processing in primates. Representational geometries of the last three layers of the model have characteristics similar to those observed in monkey face patches (posterior, middle and anterior patches). Furthermore, several face-processing-related phenomena reported in the literature automatically emerge as properties of this model. The representations are evolved through several computational layers, using biologically plausible learning rules. The model satisfies face inversion effect, composite face effect, other race effect, view and identity selectivity, and canonical face views. To our knowledge, no models have so far been proposed with this performance and agreement with biological data.

## 2 Introduction

Face recognition is robustly performed by human and non-human primates despite many transformations in size, position, and viewpoint of faces. The mechanism of face processing has been extensively studied in different modalities and species (e.g. Perrett et al., 1992; Tsao et al., 2003; Moeller et al., 2008; Freiwald and Tsao, 2010; Kanwisher et al., 1997a; Grill-Spector et al., 2004; Tsao et al., 2006; Tsao and Livingstone, 2008; McMahon et al., 2014), indicating its crucial role in understating many aspects of the cognitive processes in the brain. Electrophysiological and functional imaging studies have shown that faces are processed in specialized networks in primate's brain (Freiwald and Tsao, 2010; Grill-Spector et al., 2004; Kanwisher et al., 1997a; Moeller et al., 2008), meaning that a particular mechanism is involved in face processing. In addition, there are several face-specific perceptual phenomena such as Composite Face Effect (CFE) (Laguesse et al., 2013; Rossion, 2013; Rossion and Boremanse, 2008), Inversion Effect (IE) (Riesenhuber et al., 2004; Rossion, 2008; Rossion and Boremanse, 2008, 2008; De Heering et al., 2012), and Other-Race Effect (ORE) (Michel et al., 2006; Mondloch et al., 2010; Rossion and Michel, 2011), only applicable to face images.

Functional Magnetic Resonance Imaging (fMRI) on monkeys' brain has revealed six discrete face-selective regions, consisting of one posterior face patch [posterior lateral (PL)], two middle face patches [middle lateral (ML) and middle fundus (MF)], and three anterior face patches [anterior fundus (AF), anterior lateral (AL), and anterior medial (AM)], spanning the entire extent of the temporal lobe (Moeller et al., 2008). Each region has a different role in face processing. Cell recording from neurons in these areas of monkey brain suggests a functionally hierarchical organization for face processing in monkeys (Freiwald and Tsao, 2010). First in the hierarchy is PL, which contains a high concentration of face-selective cells, driven by the presence of face components (Issa and DiCarlo, 2012). Middle patches represent simple properties of faces (e.g. face-views) and in anterior parts, neurons become selective to more complex face properties (e.g.. face identities– Freiwald and Tsao, 2010).

There is a broad support for a general class of computational models based on the hierarchical organization of the visual pathway (reviewed in: Poggio and Serre, 2013; Poggio and Ullman, 2013; Serre, 2014; Khaligh-Razavi, 2014). These models have tried to simulate the selectivity and tolerance, which exist throughout the visual hierarchy, to stimulus transformations (Serre et al., 2007; Rajaei et al., 2012; Ghodrati et al., 2012). However, several studies have revealed that this class of hierarchical models of simple-to-complex cells, although partially successful, they fail to fully explain human object recognition mechanisms (e.g. Kriegeskorte et al., 2008a, 2008b; Khaligh-Razavi and Kriegeskorte, 2014; Ghodrati et al., 2014a). Recent modeling studies have tried to implement some face-specific properties (Leibo et al., 2011; Tan and Poggio, 2013). They have had valuable contributions in developing face processing models; and were able to explain some face-related phenomena such as, invariance and holistic face processing. However, the underlying computational mechanism of face processing and what happens in face-specific areas, such as face patches, has remained unknown. Our proposed model extends previous developments, and reaches an ideal level in which it explains neural response characteristics of monkey face patches; as well several behavioral phenomena observed in humans.

Our proposed model of face processing is based on recent electrophysiological evidence in monkey face selective areas (Freiwald and Tsao, 2010; Moeller et al., 2008; Tsao et al., 2006). The model has several layers with an organization similar to that of the hierarchical structure of the face processing system. Layers of our model simulate different aspects of face processing and its representational space similar to that of monkey face patches (Freiwald and Tsao, 2010),. The model has view selective and identity selective layers consistent with physiological and psychophysical data.

To evaluate the ability of the model in simulating representational space of faces in face-selective areas in the visual system, we compare the model responses with neuronal and some challenging behavioral data. The model is compatible with electrophysiological data for face identification, and layers of the model mimic the representational space of ML/MF and AM patches in monkeys. It is also consistent with the psychophysical data supporting the phenomena of canonical face view. The idea of canonical face view refers to the observation that specific face views carry a higher amount of information about face identities, therefore face identification performance for these views is significantly higher (Blanz et al., 1999; Liu and Chaudhuri, 2002; O’toole et al., 1998). Similarly, the model shows more invariance properties around specific views, i.e. canonical views (Blanz et al., 1999). Furthermore, we also show that the model resembles the composite face effect (CFE), a well-studied perceptual phenomenon that affects face identification in humans. Many studies have used this effect to illustrate that face perception is performed through integration of different face parts as a whole, indicating that the visual system processes faces holistically (Rossion, 2014, 2013). We also tested the model in an experiment designed to study the face inversion effect, in which upside-down inverted face images are presented to the model. Subjects' performances in face identification drop significantly when inverted faces are presented (Yin, 1969), a widely used stimulus manipulation to investigate face recognition (Jiang et al., 2006).

Humans are better at recognizing faces of own race than other races (ORE); this is another well-studied effect in the face literature (Michel et al., 2006; Mondloch et al., 2010). We also studied this effect in the model; the model nicely explains psychophysical data in two challenging databases[1-2].

Taken together, the results of multiple experiments and comparisons suggest that the proposed model very well explains the available cell recording and behavioral data. Although the proposed model is inspired by the organization of face patches in monkeys, it has the ability to explain several human behavioral face specific phenomena. This also supports the idea that man and monkey share many properties for face and object recognition (e.g. Cichy et al., 2014; Kriegeskorte et al., 2008).

# 3 Materials and Methods

## 3.1 Model Overview

The proposed model has a hierarchical structure with 6 processing layers, agreeing well with the hierarchy of the ventral visual pathway and face patches in monkey's brain (starting from posterior area (PL) to middle parts (ML/MF) and extending to the most anterior part AM). The first four layers of the model extract primary visual features, such as edges and more complex visual patterns. These layers are similar to the first four layers of the HMAX model, a biologically plausible model for object recognition (Serre et al., 2007). The last two layers of the proposed model simulate middle (ML/MF) and anterior (AM) face patches in monkey IT cortex, consistent with electrophysiological data in Freiwald and Tsao (2010) and Moeller et al. (2008). These two layers are called view selective layer (VSL-simulating middle patches, ML/MF) and identity selective layer (ISL-simulating AM). Figure 1.A schematically shows the properties of each layer. Figures 1.B and 1.C indicate the learning procedure during training and evaluation phases. Figure 1.B shows the number of subjects selected during the learning procedure across different trials. As shown in the color-coded pattern, more units are added to the ISL at the beginning of the learning procedure compared to later stages where number of face identities presented to the model is increased. Identification performance and View-invariant Identity Selectivity Index (VISI) –VISI is described in section 3.5– were used as the criteria to decide whether new units should be added to the model, Figure 1.C.

[1] www.faceplace. org; stimulus images courtesy of Michael J. Tarr, Center for the Neural Basis of Cognition, Carnegie Mellon University

[2] http://robotics.csie.ncku.edu.tw/Database.htm

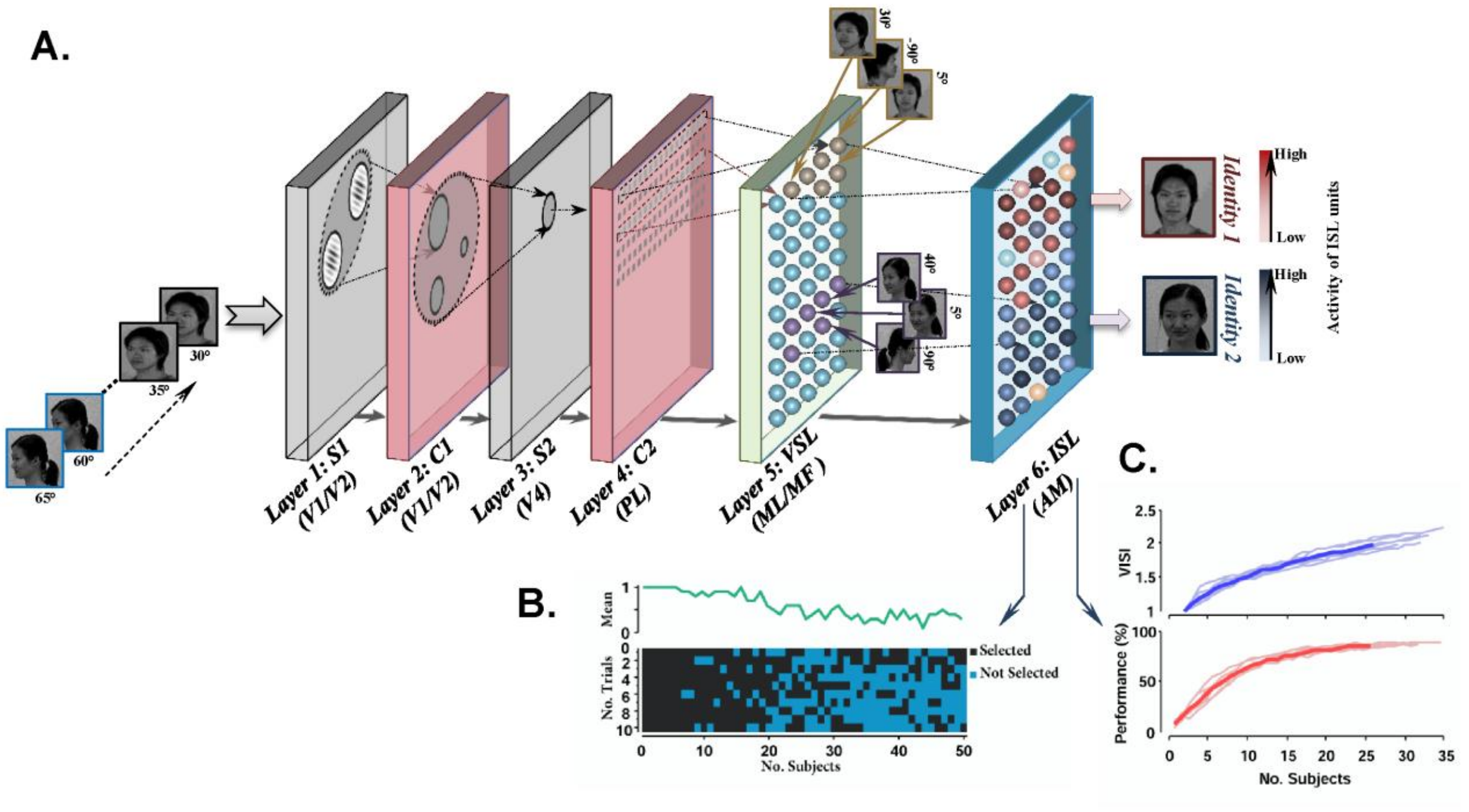

**Figure 1. Schematic of the proposed model A.** Each block shows a layer of the model with their properties. S1 and C1 layers represent bars and edges similar to V1/V2 in the visual system. Face parts are represented through S2 and C2 layers. Subsequently, face views are coded in VSL and face identities are coded within the pattern of activities in ISL units (e.g. red circles for Identity 1 and blue circles for identity 2– different shades of red/blue indicate the level of activity). **B.** Number of selected subjects in ISL during learning: The horizontal axis shows the number of ISL units (No. Subjects) and the vertical axis depicts the number of trials. The green curve shows the average of selected units across 10 random trials. **C.** VISI and identification performance saturation during learning: The horizontal axis depicts the number of selected ISL units (No. Subjects) and the vertical axis illustrates performance and VISI. The pale curves indicate 10 random runs and the thick (blue and red) curves indicate the average.

### 3.1.1 Primary feature extraction layers (S1, C1, S2, and C2)

The first two consecutive layers, S1 and C1, simulate simple and complex cells in the early visual cortex. S1 units are tuned to oriented bars within their receptive field (RF), similar to simple cells in the visual cortex (Hubel and Wiesel, 1974, 1963; LeVay et al., 1975). C1 units create slight invariant responses to scale and position of their preferred stimuli using a local max pooling operation over S1 units of the same orientation but different positions and scales (Serre et al., 2007)

The subsequent layer is S2. Units in this layer receive their inputs from small portions in C1 responses. The units are selective to the particular prototypes that are randomly extracted from training images in the learning phase. Each prototype is set as the preferred stimulus of a neuron/unit in the S2 layer, the more similar the input image to the prototype, the stronger the responses generated in S2 units. Each prototype is set as the center of a Gaussian-like function in which the distance of input image is calculated relative to the center, equation 1:

$$R = e^{-\gamma \|I - P_i\|^2} \qquad (1)$$

Where $R$ is the output response, $\gamma$ is the sharpness of the tuning function, $I$ is the input image and $P$ is the extracted prototype. We implemented 1,000 S2 units.

Each unit in the next layer, C2, performs a global max pooling over S2 units with the same prototype in various positions and scales. C2 output is a feature vector for every input image, elements of which explain the degree of similarity between prototypes and the input image. S2 and C2 units have larger receptive fields and are selective to more complex patterns than simple bars and edges. These layers simulate the responses of V4 and anterior IT neurons (PL in monkey cortex).

### 3.1.2 View Selective Layer (VSL)

Units in the view selective layer (VSL) receive their inputs from C2 layer through Gaussian tuning functions. Each unit in VSL responds to a specific pattern of C2 responses. For example, for any input image a vector of C2 values (i.e. 1000 C2 features) is compared with a set of vectors that are centers of Gaussian functions in the view selective units. These centers are tuned during the learning phase to different face views (see *Learning procedure*). In this way, different face views are represented over a population of VSL units. Each input image, from evaluation and test dataset, is represented over VSL units, using approximately 300 units (this number may change depending on the learning). The tuning properties in VSL units is inspired by ML/MF neurons in monkey face patches, which are selective to the face view (Freiwald and Tsao, 2010).

### 3.1.3 Identity Selective Layer (ISL)

Units in the identity selective layer (ISL) pool inputs through max operation, increasing invariance to the face views. Components in this layer receive connections from several VSL units with different view selectivity. The connections between VSL and ISL units are built up in the learning phase (described in the next section: *Learning procedure*). This is done by correlating face views of the same identity across time (temporal correlation); the idea being that in the real world, face views of an identity smoothly changes in time (abrupt changes of view are not expected). The time interval between face views of two identities (sequence of showing two identities) causes VSL units to make connections with different ISL units. Thus, VSL units with the same identity should be connected to one ISL unit.

View independent identity information is coded in a population of neurons in the AM face patch in monkeys. Consistently, face identities and views of novel subjects create a specific pattern of activities in the ISL units (less than 50 units in our experiments is created during the learning procedure), making a representational space for different face identities.

## 3.2 Learning procedure

Learning occurs throughout areas in the visual system, especially in higher order areas (e.g. Merzenich and Sameshima, 1993; Gilbert, 1996; Kourtzi and DiCarlo, 2006; Gilbert and Li, 2012). Likewise, computational models adapt the wiring of layers to the statistics of input stimuli using learning mechanisms. In our proposed model of face processing, learning occurs in three layers: S2, VSL, and ISL. S2 layer simply learns a dictionary of prototypes (face parts); learning in the next two layers is based on a modified trace rule (in VSL) and a continuous invariant learning (in ISL).

Learning starts with tuning of S2 units (with prototypes of four sizes: 4, 8, 12, 16), using an unsupervised random selection mechanism from training images. In the next steps, the model uses a combination of two learning mechanisms: a modified trace rule (Isik et al., 2012) and adaptive resonance theory (ART– Grossberg, 2013; Rajaei et al., 2012) to modify connection weights between C2 and VSL; as well as VSL and ISL.

### 3.2.1 Learning a dictionary of face parts in S2 layer

During the learning phase, each unit in the S2 layer becomes selective to face parts, while training face images are being presented to the model. In every presentation of a face image, several S2 units become tuned to the image parts that fall within their receptive fields. These parts are mostly face components such as eye, nose, mouth, and/or combinations of them (Ghodrati et al., 2014b). Responses of S2 units (1000 units) are maximal when the new input image matches the learned patterns. These units model the functional properties of neurons in the PL face patch in monkeys.

### 3.2.2 Continuous view-invariant learning rule in VSL & ISL

In everyday life, we continuously perceive various views of a person's face. Therefore, adjacent face views are continually perceived across time. Consistent with this characteristic, we proposed a learning mechanism to construct a view-invariant face identity representation in the model.

The learning occurs simultaneously in the last two layers (i.e. VSL and ISL) when the model is fed with input images. Units of VSL are trained using a trace learning rule shown in equation 2:

$$y_i^\tau = (1-\alpha)e^{-\frac{1}{2\sigma}(X-P_i)^2} + \alpha y_i^{\tau-1} \qquad y_i^\tau < \rho \qquad (2)$$

Where, $P_i$ is the $i^{\text{th}}$ template saved as the kernel of a Gaussian function, ρ is the vigilance parameter (threshold), σ defines the sharpness of the tuning, which is set to a constant value (σ=0.5) in a separate evaluation phase, and α is a coefficient that adds previous activity to the current output (α=0.3).

The term $\alpha\, y_i^{\tau-1}$ determines trace (memory) from previous responses. Constant ρ is a threshold value that the model uses to add a new unit to the layer. To find out whether the learned unit is sufficient to represent the input, it is compared with ρ that determines the degree to which the unit properly represents the input; the optimal value for ρ is set in the evaluation phase –using a non-overlapping set of stimuli used only for evaluation. If the activity of the learned unit is lower than ρ, the learned unit has a poor representation of the input; so, a new unit is added to the VSL population that represents the input. At the same time, a new connection between the VSL unit and the active ISL unit is established using modified trace rule. These connections are developed through the learning process and build the invariant face identification space. For example, different face views of an identity create almost the same pattern of activities in the ISL feature space. There is thus a particular representation for each identity that can be easily distinguished from others. The learning in the ISL is based on the learning rule shown in equation 3:

$$z_j^\tau = (1-\beta).Max(y_i^\tau) + \beta.z_j^{\tau-1}$$
$$w_{ij} = \begin{cases} 1 & y_i^\tau > \rho \\ 0 & otherwise \end{cases} \qquad (3)$$

Where $z_j^\tau$ is the response of the $j^{th}$ ISL unit at time τ, $y_i^\tau$ illustrates the activity of the previous layer. ISL function consists of two parts: (1) the initial part that applies a maximum operation to its inputs, with $1-\beta$ as the coefficient. (2) The trace part that includes previous synaptic activities, with $\beta$ as the coefficient. The connection weights (*w*) between the ISL and VSL are binary. In the learning phase, when a new VSL unit shows a significant response greater than vigilance parameter (ρ), the unit is connected to the winner ISL unit. Thus, the weight between these two units is set to 1 ($w_{ij}$=1).

### 3.3 Model evaluation

Images in the learning phase are sequentially presented to the model, 50 identities each in 37 views, starting with all views of an identity in random order and continuing to other identities. In order to avoid any learning bias to specific face views, while images are presented to the model, the first view of every identity is randomly selected and then other views (36 views) are presented in a sequential manner (e.g. if the first view is $45^\circ$, the next views are $50^\circ$, $55^\circ$, and so on).

The first image is applied to the model. Then, if there is not any unit in VSL, a unit with a Gaussian-like function that is tuned to the input stimuli is created. The second input image is subsequently presented to the model. Depending on the similarity of the input with the unit's preferred stimuli, a new unit can be added to the VSL and correspondingly a connection is formed between this unit and another unit in ISL. After presenting all images of an identity (different views) a blank gray image is presented to the model. This blank gray image does not generate any activity in the units (baseline); therefore, all ISL units become silent until the next input is presented to the model. As a result, previous activities do not affect new input images and the trace, especially in the last two layers, is removed.

After each step in the learning phase (i.e. whenever a new unit is added to the model), we have an evaluation phase to test the model discriminability between new identities. For this purpose, we use an evaluation dataset. The dataset contains 740 face images (20 identities, each in 37 views) that travel through the model's hierarchical structure and produce different patterns of activities, especially in the last layer. Finally, the discriminability between identities is measured and compared to the previous state of the model (before adding new units), using a View-invariant Identity Selectivity Index (VISI) and a support vector machine (SVM) classifier identification performance as measures of identity selectivity and invariant face recognition, respectively. The VISI value is compared with a threshold; a value less than the threshold indicates that the new modification (units added to the model) had no significant impact on improving the discriminability. Therefore, the new added units are removed. As the representational space is developed, the learning process is saturated (i.e. goes from coarse to fine), and only a few units will be added to the model, Figure 1. An SVM classifier is also trained on 18 face views of 20 identities of evaluation dataset (randomly selected from 37 face views) and tested on 19 face views. As shown in Figure 1.C the identification performance is saturated during the learning procedure. When the learning procedure finishes, the model becomes fixed and does not alter in further experiments.

### 3.4 View Selectivity Index

To calculate view selectivity index, a similarity matrix (Khaligh-Razavi and Kriegeskorte, 2014; Kriegeskorte et al., 2008a, 2008b; Nili et al., 2014) was computed from responses of three last layers. We then computed "View Selectivity Index" as follows: For each 740 × 740 similarity matrix (X) for test images (20 identities in 37 views), we computed the mean correlation along the squares (20 × 20) around the main diagonal of X and divided by the average of other parts of the matrix. The values of the main diagonal were omitted from the calculation because the correlation is always one on the diagonal.

## 3.5 View-invariant Identity Selectivity Index

To calculate view-invariant identity selectivity index, a similarity matrix (Kriegeskorte et al., 2008a, 2008b; Khaligh-Razavi and Kriegeskorte, 2014) was computed from responses of ISL units. We then computed "view-invariant identity selectivity index" as follows: For each 740 × 740 similarity matrix (*X*), we computed the mean correlation along the off-center diagonals {y=x+20, y=x+40… y=x+720} of X. View-invariant identity selectivity index was finally obtained using equation 4:

$$VISI = \frac{\sum_{i=}^{\{20,40\ldots740\}} \sum_{j=1}^{740} X\left(j, \mathrm{mod}\left(j+i-1,740\right)+1\right) / \left(36\times740\right)}{\sum_{i=}^{\{1,2\ldots739\}/\{20,40\ldots720\}} \sum_{j=1}^{740} X\left(j, \mathrm{mod}\left(j+i-1,740\right)+1\right) / \left(703\times740\right)} \quad (4)$$

## 3.6 Image data sets

To evaluate the model in different experiments, we used several face image datasets. All datasets are widely-used face image datasets that are freely available. We provide a brief description about each dataset in the following sections.

### 3.6.1 NCKU Face

We used NCKU dataset as a major face image dataset to train the proposed model since it contains face images with a precise variation in views. The database contains 3330 images of 90 subjects. There are 37 images, taken from 37 different viewing angles, for each identity. The viewing angles change from +90° (right profile) to -90° (left profile), with steps of 5°. Figure S5.A shows several sample images from the dataset. The dataset is freely available on *http://robotics.csie.ncku.edu.tw/Databases/FaceDetect_PoseEstimate.htm* (Chen and Lien, 2009)

### 3.6.2 Face Place

This face database was created by Tarr lab[3]. It has been used in experiments studying other race effect. We tested the model using the Asian and Caucasian races (similar to ORE psychophysics experiments: McGugin et al., 2011; Michel et al., 2006; Mondloch et al., 2010). This part of the database includes images from 38 individuals of two races with consistent lighting, multiple

[3] www.tarrlab.org; stimulus images courtesy of Michael J. Tarr, Center for the Neural Basis of Cognition, Carnegie Mellon University

views, and real emotions. Images of each identity come in seven views (+90°, +60°, +30°, 0°, -30°, -60°, -90°). Face images have a uniform white background. Several sample images are shown in Figure S5.C. The dataset is freely available through: http://www.tarrlab.org.

### 3.6.3 Composite face stimuli

The Composite face stimuli (Rossion, 2013) have been built with the purpose of investigating the composite face effect in psychophysical and neurophysiological studies. There are images of 10 different identities and 5 compositions per condition (aligned and misaligned), resulting in 50 different images in each condition (100 images in total). In aligned face images, the upper half of a face image of an identity is combined with five different lower halves in a normal face configuration. In the misaligned condition, there are similar combinations with aligned faces, but upper and lower halves do not make a normal face configuration. Figure S5.B demonstrates several samples of face images from this database. The dataset is freely available on http://face-categorization-lab.webnode.com/resources (Rossion, 2013).

# 4 Results

Different layers of the model were analyzed; and model responses were compared with psychophysical data in humans and cell recording data in monkeys. The model performance and its similarity to biological data were assessed using representational similarity analysis (RSA-Nili et al., 2014).

## 4.1 Representation of face views and identities in the network

Views and identities of different face images are represented over the last two layers of the network. Figure 2 shows response properties of the three last layers (C2, VSL, and ISL), visualized using multidimensional scaling (MDS), similarity matrix, and two indices of view and identity selectivity (VSI and VISI, see Materials and Methods). ISL responses show clear selectivity to identities when the model is presented with different views of an identity. Figure 2.A visualizes this effect as parallel diagonal lines shown in the similarity matrix (similarity measured as Pearson's correlation). The VSL similarity matrix (Figure 2.B) is characterized with a high similarity around the main diagonal, indicating view-specific representation, but no clear identity selectivity (parallel diagonal lines similar to ISL). Responses of VSL were highly selective for face images compared to other objects. Also, different populations of neurons represent different face views (Supplementary Figure S4). A moderate degree of view-specific responses can also be seen in the activities of C2 layer, like VSL, with no selectivity for identities (Figure 2.C). MDS is a visualization method, which transforms data from a high dimensional space to a lower dimensional space (Kruskal and Wish, 1978; Shepard, 1980) . The MDS plot (Figure 2.D) shows that each identity is clustered together in ISL (for 10 sample subjects, the numbers inside the discs shows identities and different colors are used for different views). On the other hand, each cluster in VSL (Figure 2.E, different colors) represents a face view while identities are intermixed. In contrast, in the C2 space (Figure 2.F), views and identities are densely distributed and highly overlapped with each other, meaning that C2 responses are not sufficiently informative about views and identities. Similar results can be seen in the plots of VSI (View Selectivity Index) and VISI (View-invariant Identity Selectivity Index), as two quantitative indexes for the representations, Figure 1.G. Overall, C2 shows a slight selectivity for face features whereas VSL and ISL demonstrate view selectivity and identity selectivity, respectively. The response properties of three last hierarchically organized layers of the model highly resemble the responses of face patches in monkeys' IT cortex –from posterior to middle and anterior face patches (for example see: Figure 4, Supplementary Figure S7, and S8 in: Freiwald and Tsao, 2010).

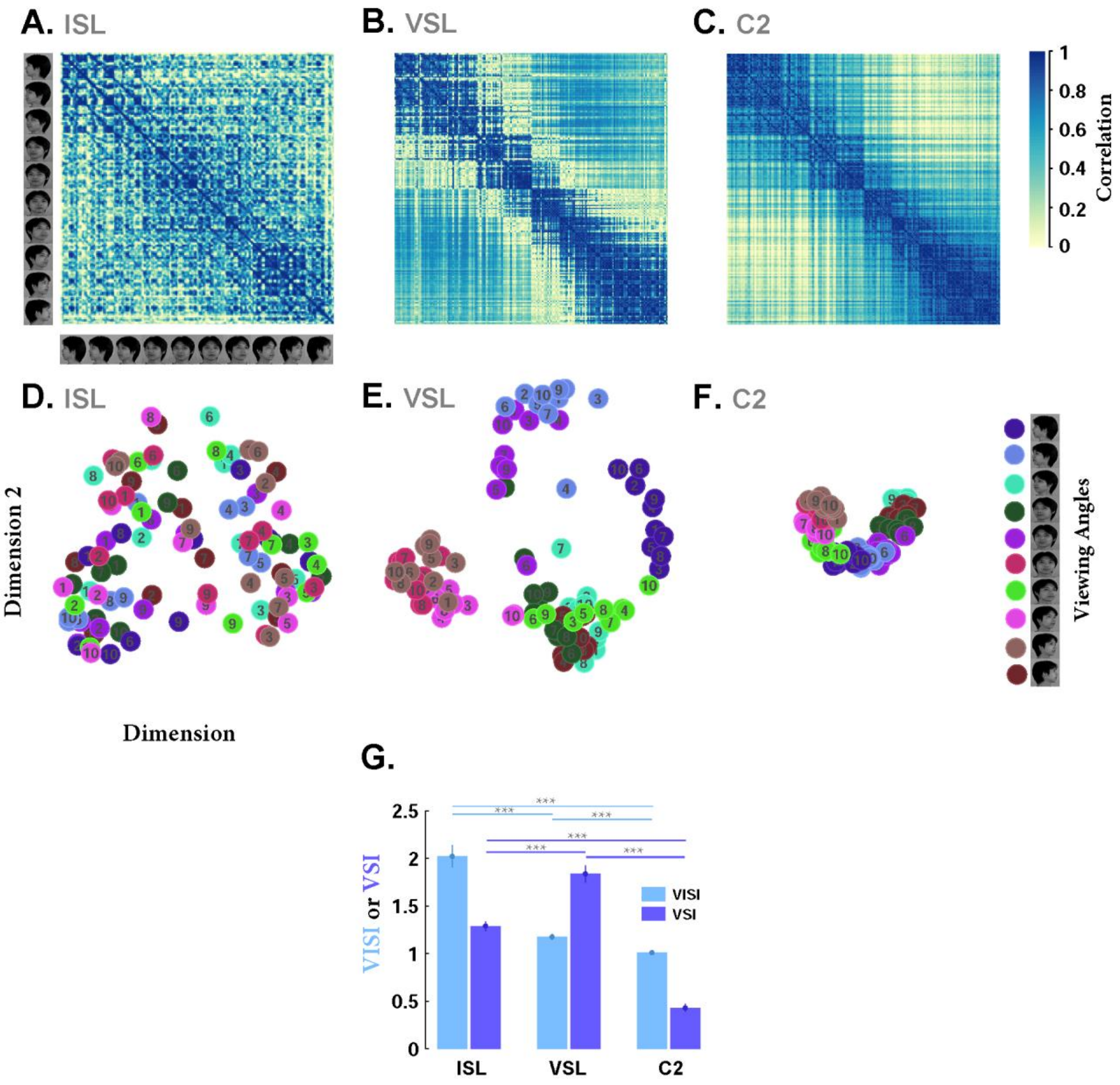

**Figure 2. Representational geometries of face views and identities in ISL, VSL, and C2.** Top row (A to C). Similarity matrices computed based on activities in ISL, VSL, and C2, from left to right, respectively. A 200×200 matrix of correlation coefficients (Pearson's correlation) was computed between feature vectors corresponding to a layer output for 10 sample subjects (face views are in the steps of 20° from -90° to 90°). Each element in a matrix shows the pairwise similarities between the internal representations of the model's layers for a pair of face views (see Material and Methods). Bottom row (D to F). Each panel depicts the results of multidimensional scaling (MDS) for responses to the face images in different layers (D: ISL, E: VSL, and F: C2). Each plot shows the location of 10 subjects (indicated by numbers from 1 to 10) at 10 face views (indicated by 10 different colors, shown in the right inset) for the first two dimensions of the MDS space. Note the clusters of the face views and identities formed in the VSL and ISL, respectively. G. VISI is significantly higher in ISL compared to VSL and C2 (ranksum test, p=0.001). Face views are better decoded in VSL compared to ISL and C2 layers.

## 4.2 Invariance to face views

Behavioral studies have shown that canonical face view, a face view between frontal and profile views, have the highest information about the face identity (Blanz et al., 1999). We investigated whether a particular face view (Blanz et al., 1999; Liu and Chaudhuri, 2002; O'toole et al., 1998) has a higher recognition performance compared to the other face views (such as full-face or profile). To this end, we used correlation analyses (Figure 3) as well as identification performances (Figure 4).

Figure 3 shows the comparison between responses of C2 and ISL units in terms of degree of invariance (DOI). C2 and ISL responses are quite different in their DOI value. To evaluate the invariance properties of the ISL and C2 features, we used a methodology similar to Logothetis and Pauls, (1995; see also: Crouzet and Serre, 2011; Pinto et al., 2011; Pinto and Adviser-Dicarlo, 2010; Riesenhuber and Poggio, 1999). View invariance was measured by first estimating a "tuning curve", obtained by correlating a feature vector corresponding to one face image at a given view with a feature vector for the same subject at different views (37 face views with the steps of 5° from -90° to 90°). An average tuning curve was then obtained by averaging similarities across views of subjects and over 10 random runs, 20 sample identities for each run. The level of invariance for each face view was determined by computing its correlation with other views of the same identity, and then averaging across the correlations; if the average was significantly higher than a pre-defined threshold, then that view has an invariant representation. The threshold is calculated for each face view by computing the maximum correlation between the feature vectors of all subjects at the same view. A tuning curve was calculated for each face view based on the activities of C2 and ISL (37 views, 37 curves–see Figure S1), representing the degree of invariance for these layers. Several samples of tuning curves are shown in Figure 3. The invariance matrices (Figure 3.A and 3.B) show the regions in which the correlation between views is significantly higher than the invariance threshold, meaning that those views carry a higher amount of view-invariant information of an identity. Consistent with behavioral studies (Blanz et al., 1999; Liu and Chaudhuri, 2002), we see a high degree of invariance in canonical views (Figure 3.C). Interestingly, this effect is more dominant in ISL compared to C2 (Figure 3.C). The DOI of ISL features is significantly higher than C2 features across all face views (Figure 3.D– $p<10^{-12}$, ranksum test).

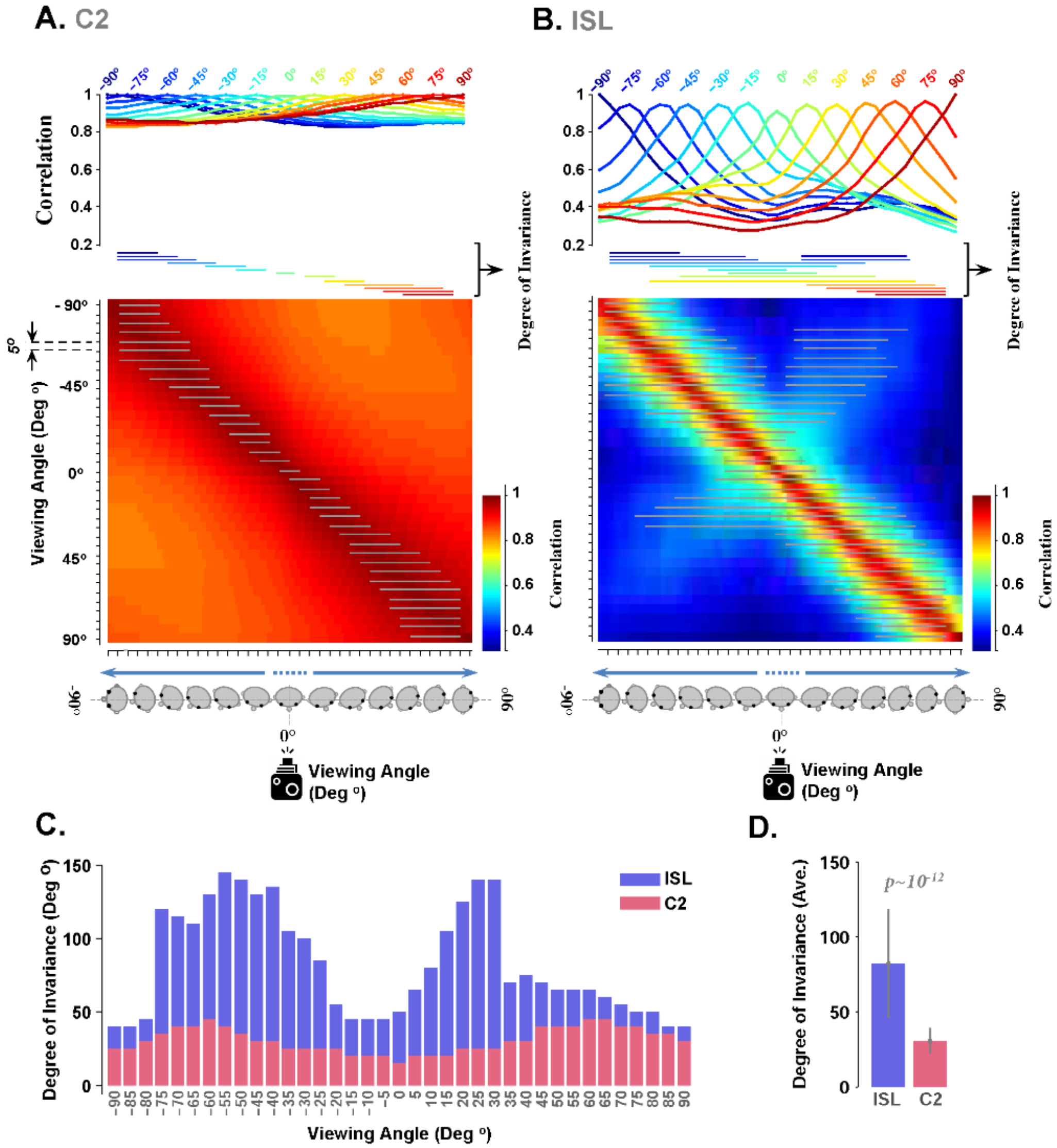

**Figure 3. Higher degree of invariance (DOI) in ISL compared to C2. A.** View invariance at the level of C2 units. Each tuning curve shows the degree of invariance in the responses of C2 units for a particular viewing angle (face view). Only a subset of tuning curves is presented (details for every view is shown in Supplementary Figure S1). The vertical axis is the correlation between feature vectors at one reference view from a set of subjects and feature vectors, computed for the same subjects across different view. The horizontal axis indicates different views with the steps of 5º. The colored, horizontal lines underneath each curve demonstrate the significant range of DOI ($p<0.02$– ranksum test) for a particular view. Each row in the invariance matrix, below the tuning curves, corresponds to a tuning curve for a face viewpoint (viewing angles are separated by 5º, from -90º in the first row to +90º in the last row. Head poses and camera position are schematically shown along the horizontal axis). Color bar at right inset represents the range of correlation. The gray horizontal lines, printed on the invariance matrix, exhibit the degree of invariance for every view similar to tuning curves (ranksum test). **B.** View invariance at the level of ISLs. **C.** Summary of view invariance responses for each face view in C2 units and ISLs. Each bar exhibits the DOI for a face view for C2 units (red bars) and ISLs (blue bars). The horizontal axis shows different face views. **D.** Average DOI across all views for ISL and C2, calculated using data shown in C.

We also analyzed the performance of the model in invariant face recognition using ISL features using support vector machine (SVM) classifiers, Figure 4. The SVM was trained with one view and tested by other views (repeated across 10 individual runs for every view, separately). The performance decreases as the views deviate from the training view, Figure 4.A (see Supplementary Figure S2 for more details). This observation might not be surprising; but, the interesting point is that the degree of invariance in ISL features increases around canonical face views, Figure 4.B. These evaluations exhibit that the model is able to represent the effect of canonical face views (views that contain more invariant information) (Blanz et al., 1999; Liu and Chaudhuri, 2002; O'toole et al., 1998).

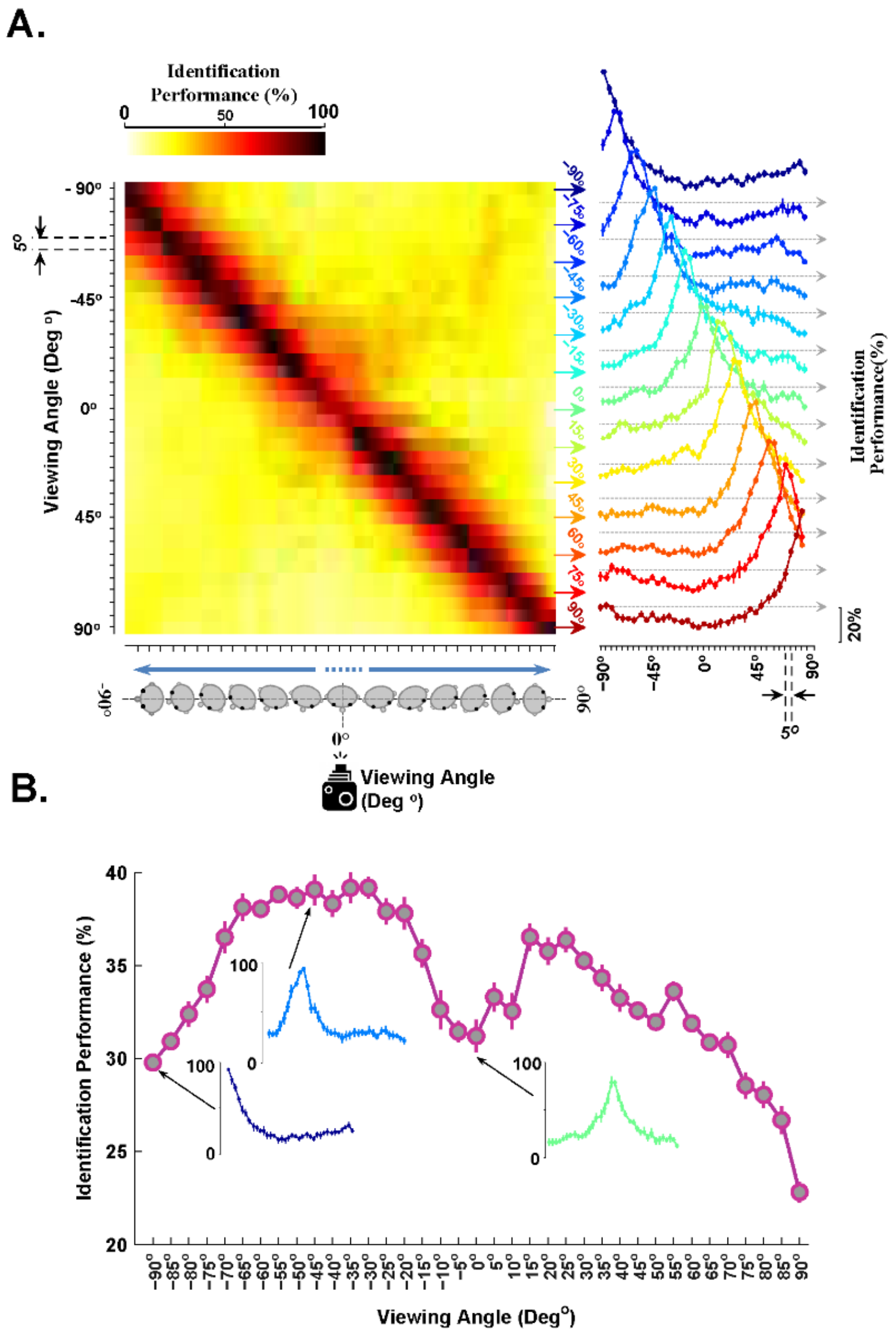

**Figure 4. Performance of the model (ISL) in view invariant face recognition. A.** The performance of face identification in different views. The color-coded matrix shows the performance of the model in identifying subjects across different views. Each row of the performance matrix illustrates the performance of the model for one view (trained using a particular view and tested over all views). The color bar at the top-left shows the range of identification performance. The vertical axis shows different face views for training. The horizontal axis corresponds to different test views, the first row of the matrix shows that a classifier trained by -90º and tested with all other views. The chance level is 5%. A subset of performance curves is shown at the right inset, demonstrating the performance variations in different views, the peaks of performance curves change as the training views change (details of performances in every view are shown in Supplementary Figure S2). The small, black, vertical axes at the right of the curves show 20% performance. Error bars are standard deviations over 10 runs. **B.** Performance comparison across different views. Each circle refers to the average of recognition rate in each view (i.e. the mean performance

across all views). The vertical axis indicates the mean performance and the horizontal axis shows different views. Several performance curves are shown for some sample views. Error bars are the standard deviation and the performances are the average of 10 runs.

## 4.3 Only ISL feature space has a dominant face inversion effect

Face inversion effect (FIE) has thoroughly been studied both physiologically and psychophysically (e.g. De Heering et al., 2012; Kanwisher et al., 1998, 1997b; Riesenhuber et al., 2004; Rossion, 2008; Rossion et al., 1999; Yovel and Kanwisher, 2005). Subjects' performances in face identification drop significantly when inverted faces are presented (Yin, 1969). This effect is one of the widely-used stimulus manipulations to investigate face recognition mechanisms in the brain. Here, we evaluated responses of the model in face identification tasks when face images were either inverted or normal. We examined the inversion effect in two layers of the model: C2 and ISL. Figure 5 shows a clear inversion effect in ISL units; however, C2 features either do not show the face inversion effect or only show a very weak effect in a few face views. We used four different approaches to investigate the inversion effect across layers of the model: measuring Euclidean distance between upright and inverted faces, computing similarity matrices across all views for both upright and inverted faces, MDS plots, and VISI. Average Euclidean distance between ISL feature vectors (averaged across all views of the same identity) is significantly higher for upright face images compared to inverted faces. Once inverted faces were fed to the model, the discriminability of the units dropped and identities seemed to be similar; therefore, the distance between representations is reduced. However, C2 features had no significant difference in their Euclidean distance for the two cases (upright/inverted), Figure 5.A. The diagonal lines in the similarity matrices and the pattern of distributions in the MDS plots were two measurements that enabled us to better investigate the inversion effect in the upright and inverted faces, Figure 5.B, C. Parallel diagonal lines in the similarity matrices of upright faces indicate that identities (10 sample subjects) are represented better compared to the inverted case in ISL (a subset of 8 face views are shown). Colors in the MDS plot, which represent identities, are clustered more strongly in upright faces than inverted (10 sample subjects, Figure 5.B). Furthermore, using VISI (see Materials and Methods), we quantitatively showed the representations of identities in the model for upright and inverted face images, Figure 5.C. VISI is significantly higher in ISL units for upright compared to invert faces, meaning that ISL activities resemble psychophysical data in humans. In addition, VISI is significantly greater in ISL compared to the C2 units (Figure 5.C – $p<10^{-4}$, rank sum test).

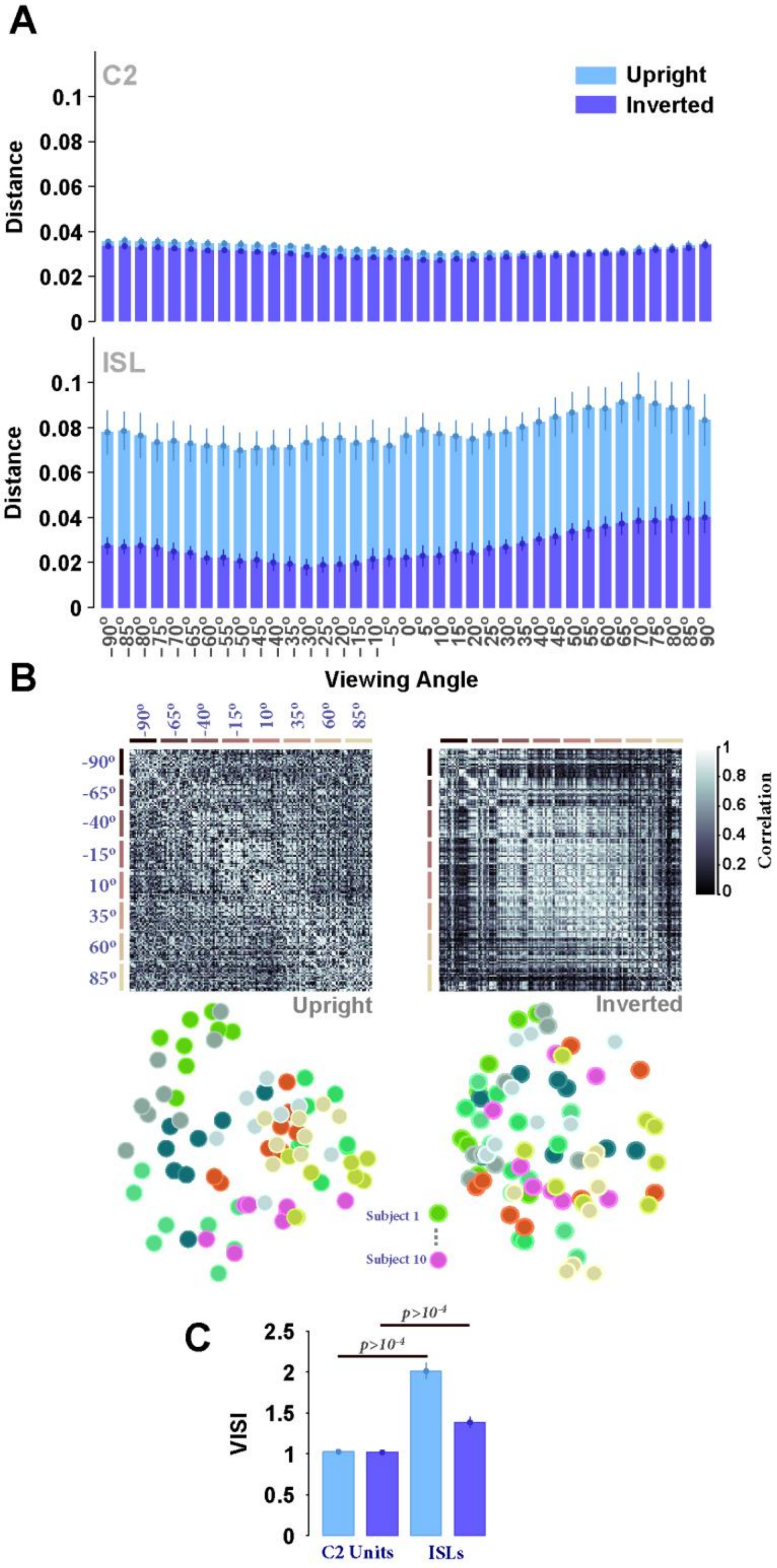

**Figure 5. Face inversion effect (FIE) for different views. A.** The distance between feature vectors of inverted and upright face images for C2 units (up) and ISL (down). Inversion effect is highly significant at ISL compared to the C2 layer (normalized Euclidean distance). The vertical axis indicates the normalized distance and the horizontal axis shows different views, separated with the steps of 5º. The cyan bars represent the results for upright face images and the purple bars show the results for inverted face images. **B.** MDS similarity matrices in the ISL upright (left) and inverted (right) faces. Similarity matrices show the pairwise similarities between the internal representations of the model for two different face views. The diagonal, parallel lines in the similarity matrix for upright faces (left) indicate the identity selectivity in the ISL for upright faces. The similarity matrix for inverted face images is shown at right. The lines along horizontal and vertical axis indicate different face views. Left MDS shows the results for upright faces while the right MDS represents the results for inverted faces. Color-coded circles in the MDS space represent subjects (10 subjects) at eight different views. **C.** VISI for upright and inverted faces in the model (ranksum test-see Materials and Methods). Error bars are the standard deviation (STD) obtained over 10 independent runs.

## 4.4 Composite face effect in ISL

One of the interesting phenomena in face perception is Composite Face Effect (CFE). When two identical top halves of a face image are aligned with different bottom halves, they are perceived as different identities, showing that face identification needs a whole face image (Rossion, 2013). Many studies have used this paradigm to illustrate that face perception is performed through the integration of different face parts as a whole, suggesting that the visual system processes faces holistically (Laguesse et al., 2013; Rossion, 2013; Rossion and Boremanse, 2008).

To investigate CFE in the proposed model, we trained the model using NCKU dataset (Chen and Lien, 2009)- see Material and Methods for details). In the test phase, the model was presented with composite face stimuli from Rossion (2013), consisting of aligned/misaligned face images of 10 identities, each having five compositions (Supplementary Figure S5). To assess model performance in a task using composite face stimuli, we measured the similarity between same top halves (different bottom halves), within 15 trials of permutation for each identity. The similarity was then compared with a threshold. If the similarity was below the threshold, two images were considered as the same identity. By changing the threshold (from >0 to <1), we calculated the hit rate in response to the aligned and misaligned cases and plotted the hit rate against threshold values. As opposed to the C2 layer, ISL responses are not significantly affected by changing the threshold value (Figure 6.B) –in C2 the hit rate significantly drops by increasing the threshold. For ISL, the hit rate in misaligned images (red curve) is significantly higher than the aligned faces (blue curve) for all thresholds above 0.25. This indicates that two identical top halves with misalignment are assumed more similar than the aligned case (i.e. having two identical top halves with aligned lower parts, which makes them to be perceived as different identities). There is no clear difference in C2 responses between aligned and misaligned faces, Figure 6.A. This suggests that face features in C2 are invariant to the alignment/misalignment of face images.

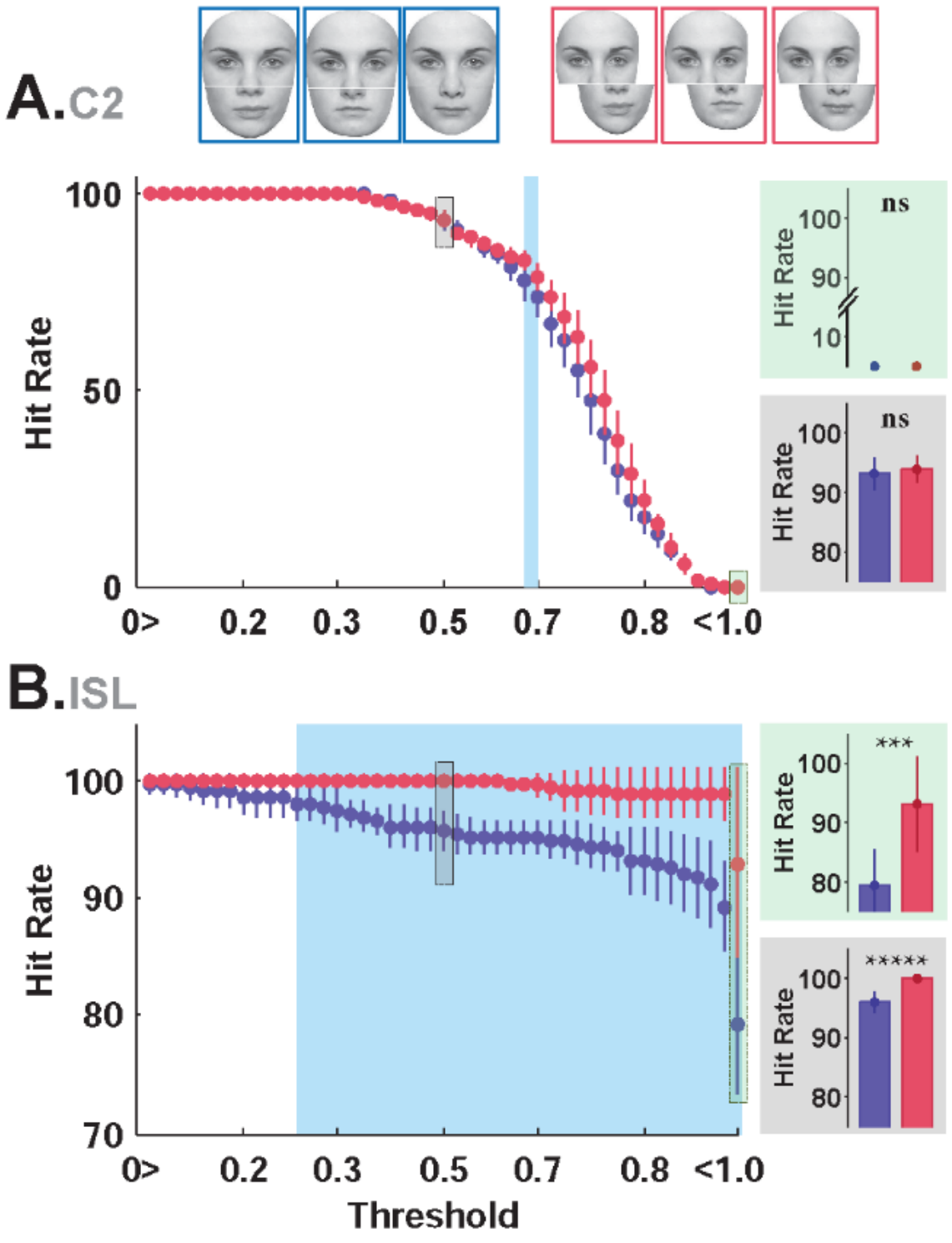

**Figure 6. Model responses in the aligned vs. misaligned face identification task (Composite Face Effect). A.** The hit rate in identification of aligned (purple) and misaligned (red) faces in the C2 layer. The vertical axis shows the hit rate while the horizontal axis shows the threshold range (see Material and Method). Several samples of aligned (purple frames) and misaligned (red frames) face images are shown at the top of the plot. Two sample bar plots are shown at the right inset for two different thresholds: 0.5 (gray background) and ~1 (green background). The blue region is the area in which the hit rate between aligned and misaligned faces is significantly different (ranksum test). **B.** The hit rate in identification of aligned and misaligned faces in ISL. In both A and B each point corresponds to the hit rate for the threshold value shown on the X-axis (different thresholds specify the boundary of the model to consider two face images as the same identity, 0<thr<1).

## 4.5 We better identify faces of our own race: Other-Race Effect (ORE)

People are better at identifying faces of their own race than other races, an effect known as other-race effect (ORE–e.g. Golby et al., 2001; McGugin et al., 2011; Michel et al., 2006; Mondloch et al., 2010). Similarly, we show in this section that the model better identifies faces of the race it is trained with.

Some studies suggest that there are different mechanisms for the identification of faces of the same and other races (e.g. holistic- versus component-based face processing– Rossion and Michel, 2011). Here, we used two face image datasets of Asian and Caucasian races to assess this effect in the proposed model and compared the responses of the model with reported data from human psychophysics (Rossion, 2013).

To test the effect, the model was first trained using Asian faces (from NCKU dataset) and tested on both Asian and Caucasian[4], Figure 7.A. Second, we investigated this effect by changing the races in the train (Caucasian) and test (Asian) phases, Figure 7.B. ORE is shown using two measures: identification performance and dissimilarity. Identification performance was measured using a SVM classifier, trained on adjacent views (-90°, -30°, 30°, 90°) and tested on middle views (-60, 0, 60), for Asian and Caucasian face images, separately. Dissimilarity was measured by computing the average Euclidean distance within the faces of the same race. In both performance and dissimilarity, the discrimination between identities is significantly higher for the same race (ranksum test– $p<0.003$; Figure 7), confirming the reported behavioral results (Michel et al., 2006; Rossion and Michel, 2011). We further investigated the effect for each of the face views separately. In ISL, almost in all views (-90°, -60°, -30°, 0°, 30°, 60°, 90°) the dissimilarity is significantly higher for the same race compared to the other race (Figure S3).

[4] www.tarrlab.org ; stimulus images courtesy of Michael J. Tarr, Center for the Neural Basis of Cognition, Carnegie Mellon University

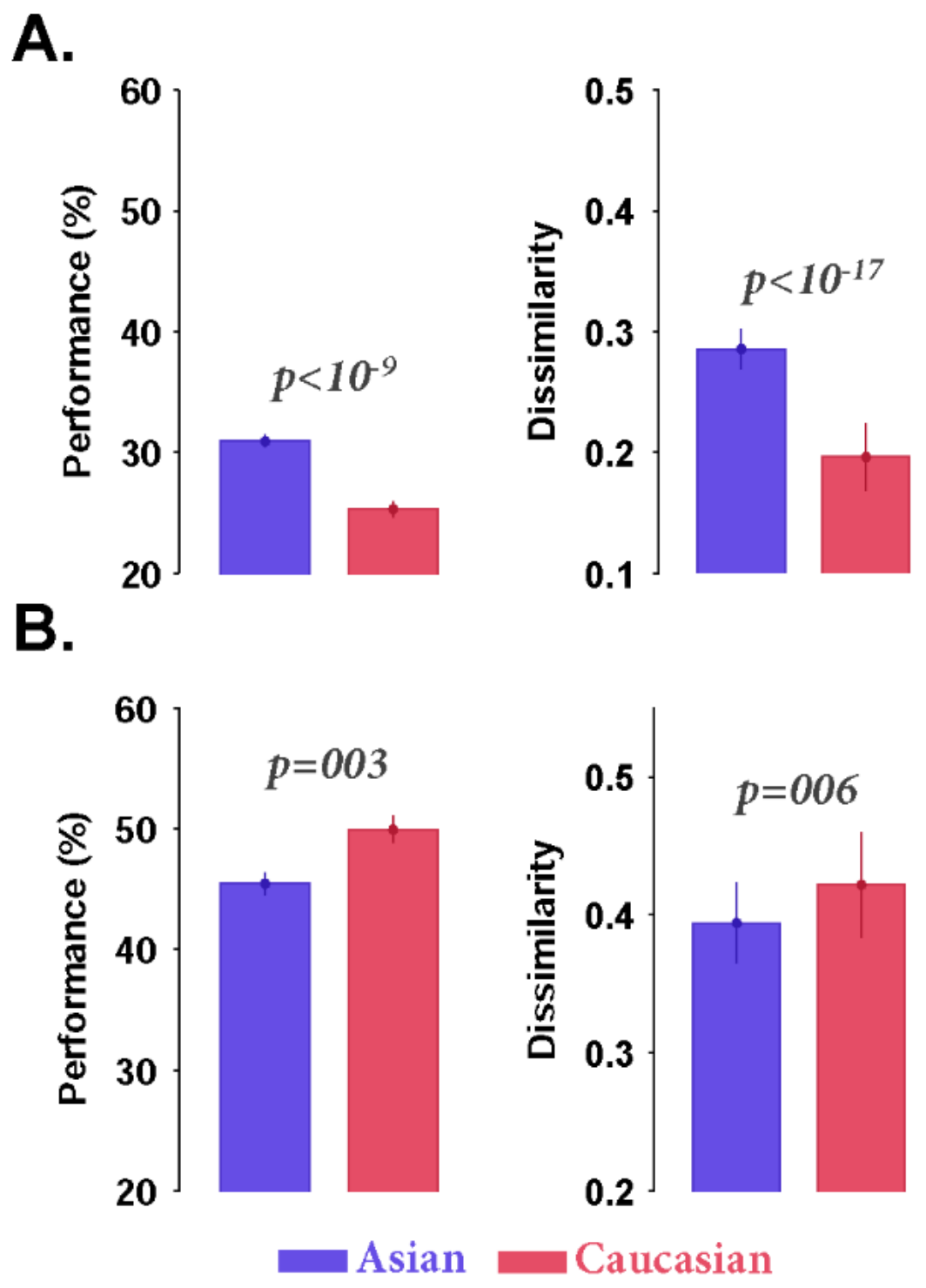

**Figure 7. Discriminability of ISL units in response to Asian and Caucasian faces. A.** The dissimilarity (right- calculated based on Euclidean distance) and performance (left) between feature vectors of different races (using ISL features). A typical other-race effect can be seen, as observed in face recognition tasks in behavioral studies. ORE is highly significant in ISL. The model was trained using images from NCKU dataset (Asian race) and tested using Asian and Caucasian images from Tarr dataset. The vertical axes indicate identification performance (left) and dissimilarity calculated based on normalized Euclidean distance (right). The blue bar indicates the results for Asian face images and the red bar shows the results for Caucasian face images. **B.** The dissimilarity (right) and performance (left) between feature vectors of different races in ISL when the model was trained on Caucasian faces and tested using both Asian and Caucasian (Tarr dataset). In all plots error bars are the standard deviation obtained over 10 runs. P-values calculated using ranksum test.

# 5 Discussion

We introduced a new biologically-plausible model for face recognition, consistent with recent cell recording evidence (Freiwald and Tsao, 2010). In particular, the model was able to account for response properties of face patches in monkeys as well as several well-studied behavioral phenomena for face processing in humans such as: face inversion effect, composite face effect, canonical face view, and other race effect. We considered both modern theories of face and object processing (i.e. population/distributed coding) and some classical, yet powerful, ideas (e.g. holistic face processing) in the model.

## 5.1 Specialized face processing

A fundamental question in biological object-vision is whether the brain utilizes the same mechanism to process all object categories or employs a specialized mechanism for particular categories (generic vs. specialized). The former is the generic view, suggesting that any object category is represented over distributed patterns of neuronal activities in the IT cortex. Objects can be discriminated based on distinctive patterns of activities (Haxby et al., 2001; Ishai et al., 2000; Schwarzlose et al., 2008; Spiridon and Kanwisher, 2002)., The latter suggests that there are specific areas in IT highly selective to some categories, such as faces (Freiwald and Tsao, 2010; Kanwisher et al., 1997b; Moeller et al., 2008; Tsao and Livingstone, 2008), scenes (Aguirre et al., 1998; Epstein and Kanwisher, 1998; Hasson et al., 2003; Maguire et al., 1998) and bodies (Downing et al., 2001; Pinsk et al., 2005). Functional MRI studies show that evoked responses in other areas, excluding face selective regions, contain sufficient information for face/non-face discrimination (Haxby et al., 2001; Spiridon and Kanwisher, 2002; Tsao et al., 2003). Therefore, face selective patches are suggested to be involved in more specific tasks of face recognition (i.e. view-invariant face identification– Freiwald and Tsao, 2010). The proposed model is an example of theories positing that face identification relies on a specific mechanism that gives the ability of finer face processing. Units of the proposed model are highly selective for face images but not for other objects (Supplementary Figure 4). This specific network enabled us to investigate several behavioral face specific phenomena.

## 5.2 Grandmother cells vs. distributed coding

The idea of grandmother cells emerged in the last two decades, indicating that there are highly selective neurons for particular objects/faces (e.g. Kreiman et al., 2000; Quiroga et al., 2005). In this coding scheme, no further processing was required to extract an object label from neuronal representations. However, it seems implausible to have a separate cell for each object because it restricts the number of objects under consideration (Bowers, 2009). Distributed coding is the other side of the debate, suggesting that the information (e.g. face identity) is distributed over a population of neurons in higher visual areas. In this scheme each neuron is involved in representation of different stimuli. Therefore, none of them needs to be precisely tuned to a particular stimulus and an extra processing stage is required to readout the representations (Bowers, 2009).

In our proposed model, face views and identities are stored over a population of several units in the model. For any given face image there are few responsive units in VSL; this is consistent with electrophysiological studies showing that face views are encoded sparsely (Freiwald and Tsao, 2010; Rolls, 2007);. Units in ISL represent face identities over a distributed patterns of activities, meaning that each unit is involved in encoding many identities and the response of a single unit is not solely informative enough about an identity. Consequently, information of an identity is distributed over responses of all units. The pattern of responses for an identity is also invariant to different views of the identity.

Face images of the same view share similar information. Since there are limited number of possible face views (e.g. 0°-360°); and the correlation between nearby views is high, coding of different views can be done using a sparse coding approach, meaning that a few number of neurons/units can represent a range of views (e.g. 0-30). The brain may also use a similar sparse coding approach for representing different face views, which is computationally efficient too.

There are infinite number of identities that need to be represented over a population of neurons in face selective areas. This requires a distributed coding approach that enables encoding many identities by eliciting different patterns of activities in face selective areas. It seems that identities are less probable to be coded using grandmother cells due to limited number of neurons that exist in face selective areas.

## 5.3 Holistic Face Processing

Several studies have suggested that faces are processed as wholes rather than individual parts, which is referred to as holistic face processing (Carey and Diamond, 1977; Farah et al., 1998; Peterson and Rhodes, 2003; Rossion and Michel, 2011; Schwarzer, 1997; Tanaka and Farah, 1993). Disturbing the configuration of face images leads to reduction in both recognition speed and accuracy (Farah et al., 1998; Tanaka and Farah, 1993). Many behavioral studies have evidenced holistic processing using different experiments (Michel et al., 2006; Rossion, 2013; Yovel and Kanwisher, 2005). We tested our model in three different well-known face experiments, supporting the idea of holistic face processing. First, we investigated the *Composite Face Effect*. When two identical top halves of a face image are aligned with different bottom halves, they are perceived as different identities and we are unable to perceive the two halves of the face separately. ISL units in our model showed a similar behaviour; the dissimilarity was higher between aligned face images than misaligned faces. This is because this layer of the model represents face images holistically and has misperception when encounters with aligned images. Second, the model shows a face *Inversion Effect*, another well-studied effect, supporting holistic face processing. Performance drops when inverted faces are presented to humans (Bruce, 1998; Maurer et al., 2002; Thompson, 1980). Upright face images are processed using configural and featural information (holistic– Tanaka and Farah, 1993), which is also regarded as evidence for multi-feature selectivity (Wallis, 2013). The discriminability of the model was reduced when inverted face images were presented. Finally, the model also showed another face-related phenomenon, known as the *Other Race Effect*, again a perceptual effect confirming holistic face processing. It is suggested that the holistic processing of face information occur for face images of our own race, which enables us to better identify individuals who have a face more similar to the average face we have as a template (Michel et al., 2006).

We showed that ISL units in the model have properties such as, composite face effect and inversion face effect, suggesting that faces are processed holistically in this layer. However, this is not an obvious feature of the C2 layer, which is considered as a part-based layer in the model analogous to PL in monkey face patches (having similar representational geometries for both upright and inverted face images, Figure 5, suggests that the C2 layer is not sensitive to holistic information such as configuration). It suggests that the C2 layer is rich enough for object recognition and face/non-face categorization, but not for face view and identity coding.

# 1 Supplementary information

**Supporting Text 1. Face Selectivity in VSL:** To evaluate the response properties of VSL units, we assessed selectivity of the units for face and object images. We selected eight diverse categories of objects (consisting of the human face, animal body, fruit, gadget, human body, animal face, plant, and scramble images; 16 images for each category) and applied them to the model. Each row of the scatter plot in Figure S4 (top) shows the mean response of a VSL unit across all (8×16) images. Color-coded values show the amount of normalized activity. As shown in Figure S4 (top), high responses of units for face-like images declare face selective responses. Furthermore, the VSL units have selectivity to specific face views,, Figure S4 (bottom). These two properties resemble response characteristics of the population of ML/MF neurons (for comparison with electrophysiological data see Figure 3 in: Freiwald and Tsao, 2010).

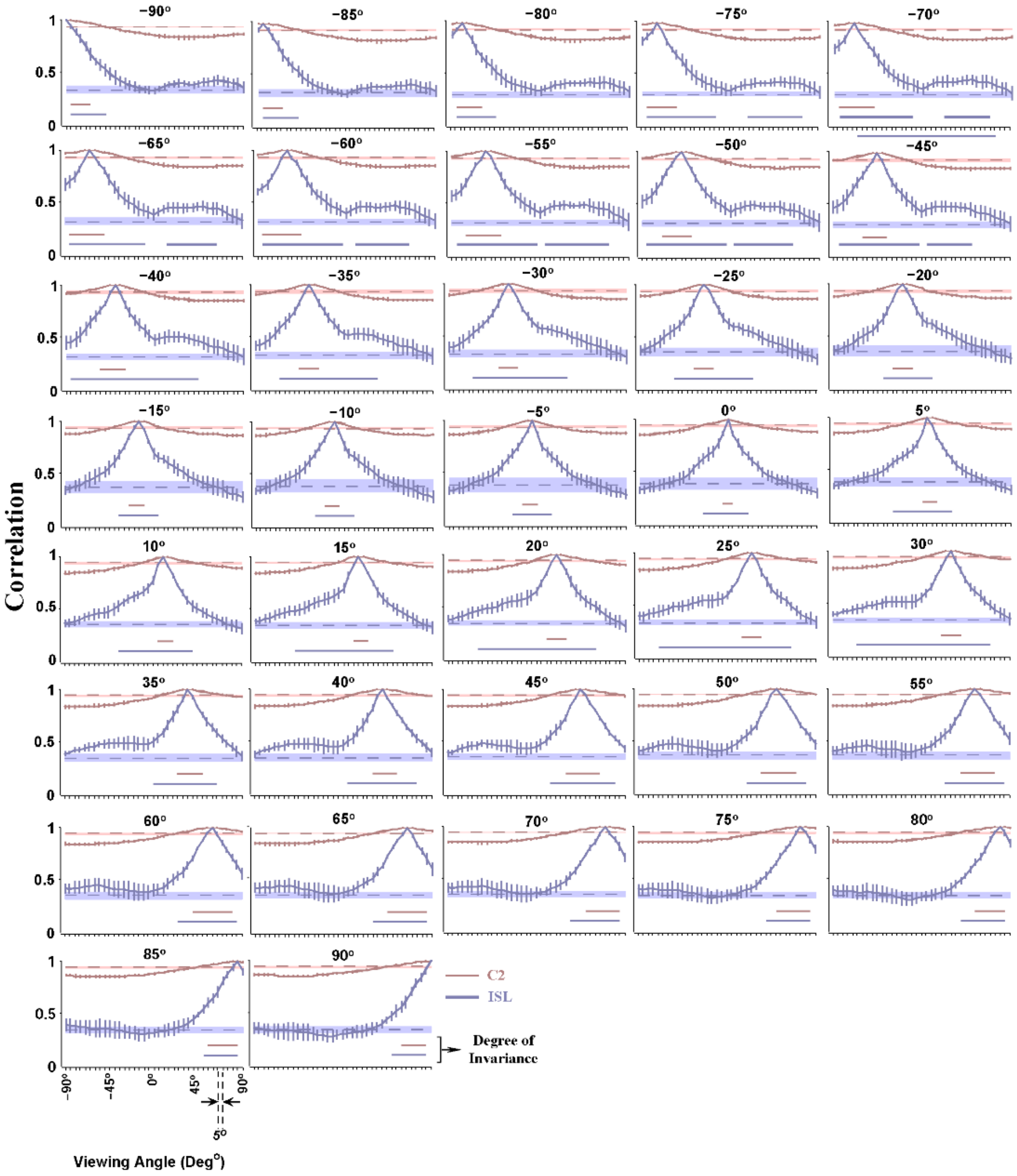

**Figure S1. View invariant tuning curves for C2 and ISL units across different face views.** The pink curves indicate tunings for C2 units and the purple curves indicate tunings for ISL units. Each point on a curve exhibits the correlation between feature vectors at one reference view from a set of subjects and feature vectors computed for the same subjects across other views. The horizontal axis shows views, separated with the steps of 5º. The horizontal dashed lines, shaded with error area, show the average correlation among feature vectors in one view of different subjects. Error bars are the standard deviation and the correlations are the average of 10 random runs. The horizontal lines, underneath the curves represent the degree of invariance for C2 responses (pink lines) and ISLs (purple lines).

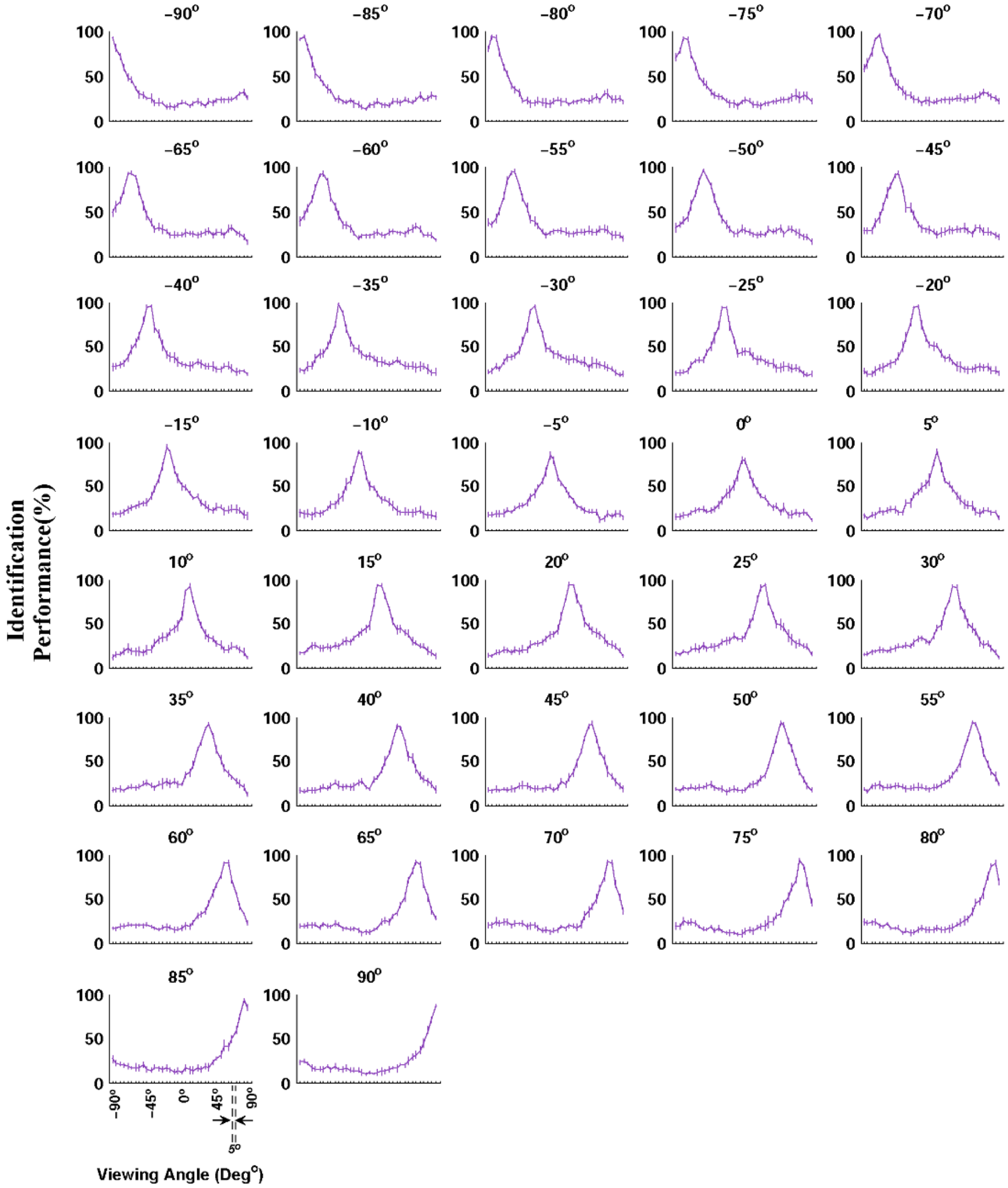

**Figure S2. Performance of the model in face identification for the ISL.** Each plot shows the performance of the model for a face view. Each point on a curve (a sample train view) shows the identification performance at one test face view from a set of subjects, 20 identities. The vertical axis indicates the identification performance and the horizontal axis shows different views separated with 5º. Error bars are standard deviations, and performances are the average of 10 runs.

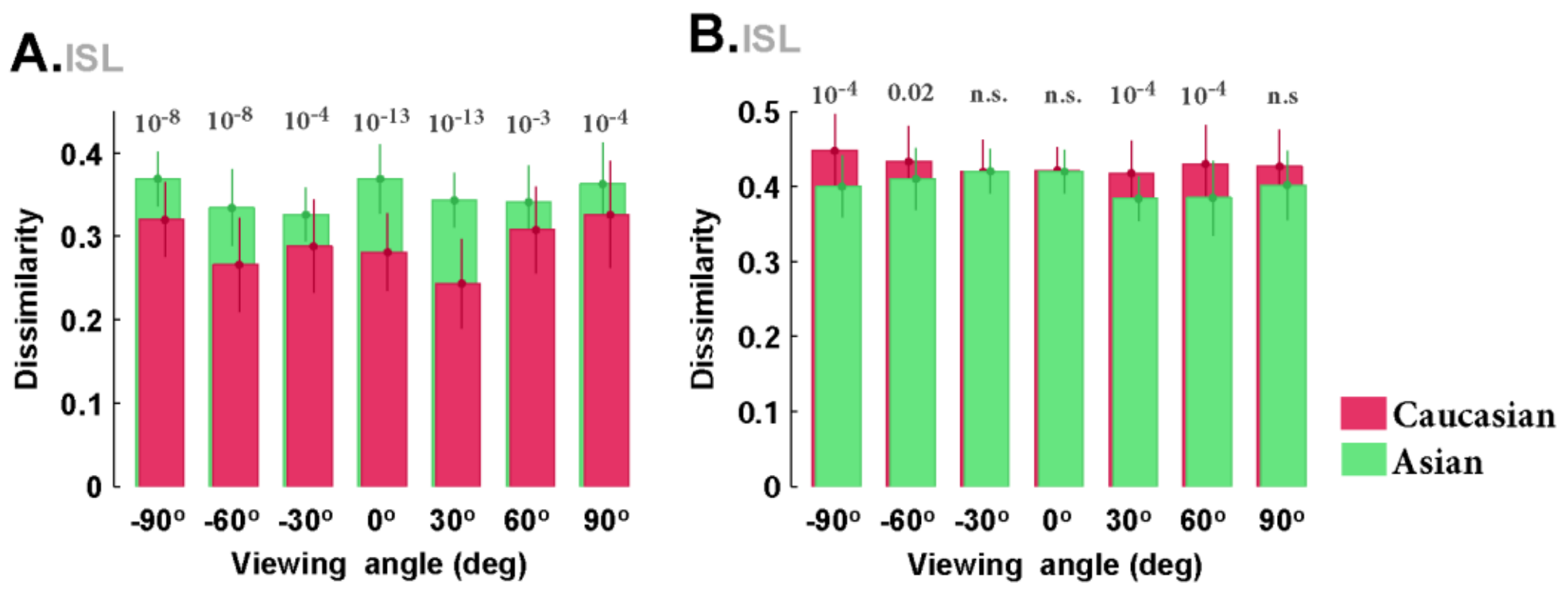

**Figure S3. Illustration of the other-race effect in ISL across different views. A.** The dissimilarity for own- vs. other-race faces as a function of face views when the model is trained by Asian and tested on both Asian and Caucasian . The horizontal axis shows seven views (+/- 90 º, +/- 60 º, +/- 30 º, 0 º) and the vertical axis indicates dissimilarity. Simulations show a decrease in dissimilarity (as a result of ORE)**. B.** The dissimilarity for own- vs. other-race faces as a function of face views when model trained on Caucasian and tested by both Asian and Caucasian in ISL. In all plots error bars indicate standard error of the mean, and dissimilarities are the average of 50 independent runs within each race. *P-values* were calculated using ranksum test.

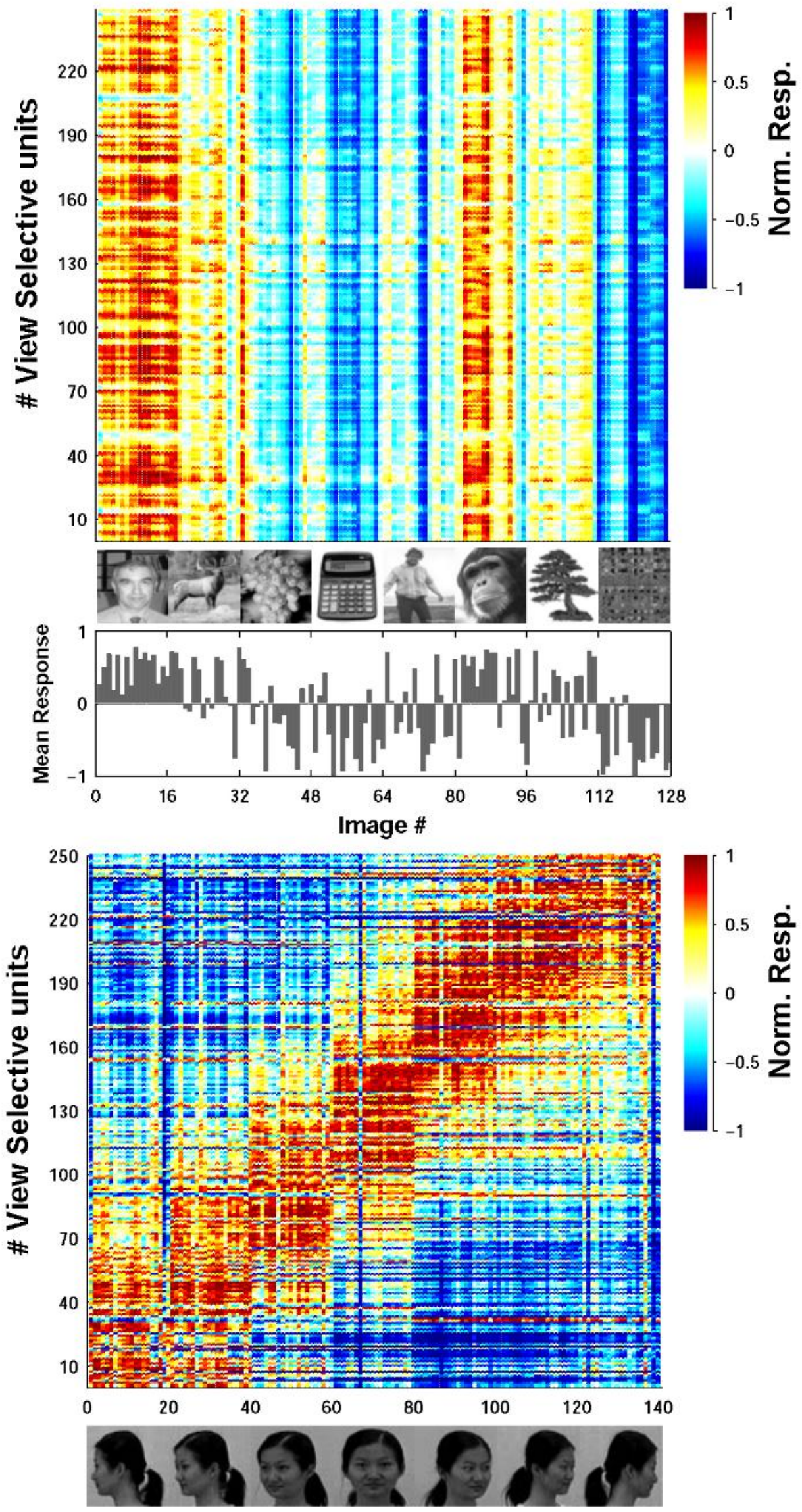

**Figure S4. Face selectivity in VSL units. Top.** VSL responses to eight categories of objects. The mean response of the units is shown in the bar plot, below the figure, the normalized activity is color-coded. The horizontal axis shows eight object classes with 16 images per class (Human face, Animal body, Fruit, Gadget, Human body, Animal Head, Plant, Scramble images) and the vertical axis indicates the normalized response of VSL units. **Bottom**. Responses of VSL units to different face views. The horizontal axis shows seven sample views for an identity. Vertical axis depicts responses of VSL units.

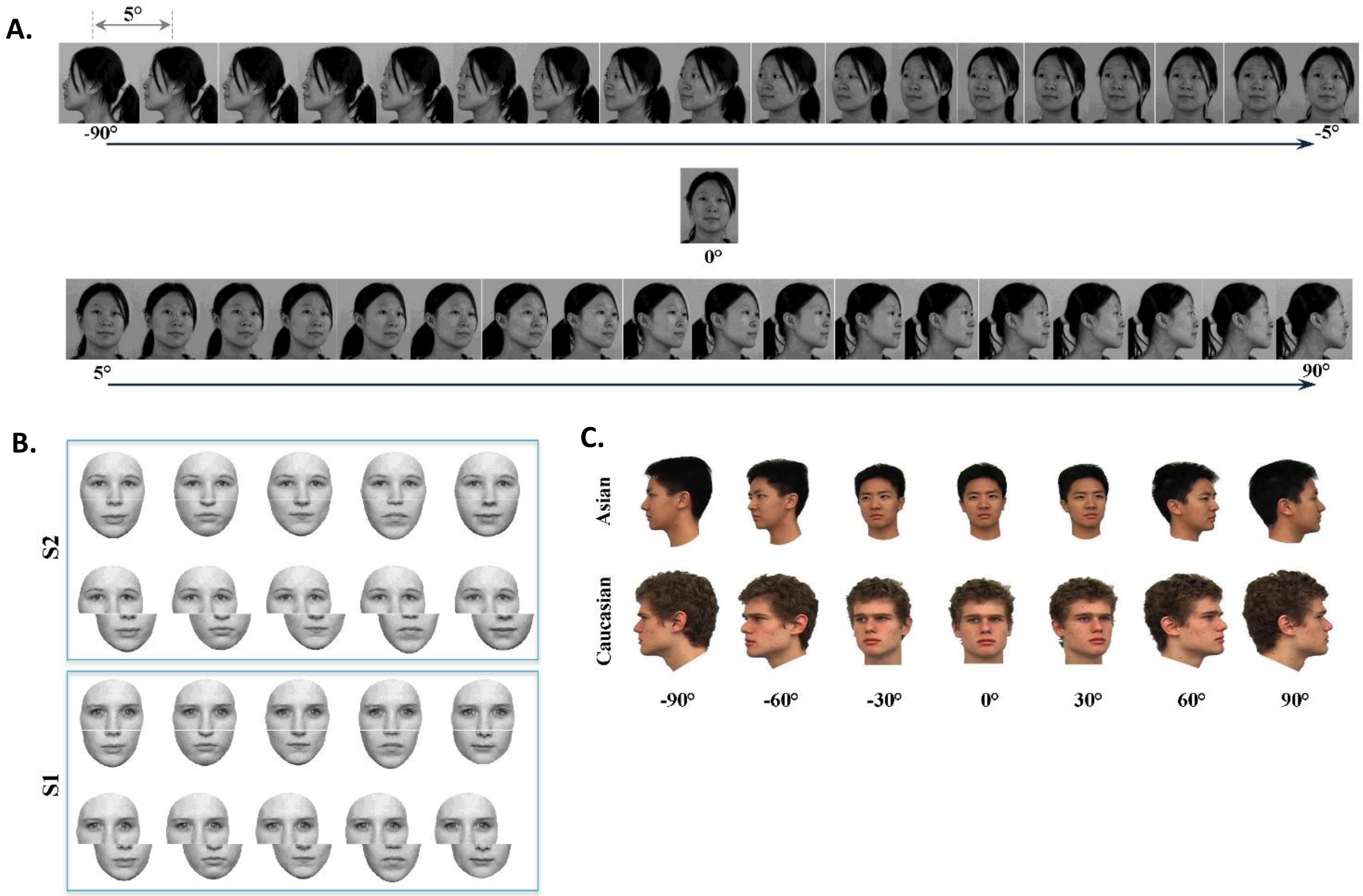

**Figure S5. Datasets A.** NCKU face dataset. The database contains 3330 images of 90 subjects. There are 37 images, taken from 37 viewing angles, for each identity. The viewing angles change from +90° (right profile) to -90° (left profile) with the steps of 5°. **B.** Composite face stimuli. There are images of 10 different identities and 5 compositions per condition (aligned and misaligned), resulting in 50 different images in each condition (100 images in total). **C.** Other race database. The model was tested using Asian and Caucasian races. This part of the database includes images from 40 individuals of two races with multiple views, and real emotions. Images of each identity come in seven views (+90°, +60°, +30°, 0°, -30°, -60°, -90°). Face images have a uniform white background.